\documentclass[twocolumn,dvipsnames]{aastex631}

\newcommand{\Msun}{\ensuremath{\rm M_{\odot}} }

\newcommand{\affuofa}{Steward Observatory, University of Arizona, 933 North Cherry Avenue, Tucson, AZ 85721, USA}

\shorttitle{Squars: Exoplanets do not Exist}
\shortauthors{WHAC Collaboration}

\graphicspath{{./}{figures/}}

\begin{document}

\title{A Modest Proposal for the Non-existence of Exoplanets:\\The Expansion of Stellar Physics to Include Squars}

\author[0000-0001-5962-7260]{Charity~Woodrum}
\affiliation{\affuofa}

\author[0000-0002-4684-9005]{Raphael~E.~Hviding}
\affiliation{\affuofa}

\author[0000-0002-1546-9763]{Rachael.~C.~Amaro}
\affiliation{\affuofa}

\author[0000-0001-8765-8670]{Katie~Chamberlain}
\affiliation{\affuofa}

\date{April 1st, 2023}

\begin{abstract}
The search for exoplanets has become a focal point of astronomical research, captivating public attention and driving scientific inquiry; however, the rush to confirm exoplanet discoveries has often overlooked potential alternative explanations leading to a scientific consensus that is overly reliant on untested assumptions and limited data.
We argue that the evidence in support of exoplanet observation is not necessarily definitive and that alternative interpretations are not only possible, but necessary.
Our conclusion is therefore concise: \underline{exoplanets do not exist}.
Here, we present the framework for a novel type of cuboid star, or \textit{squar}, which can precisely reproduce the full range of observed phenomena in stellar light curves, including the trapezoidal flux deviations (TFDs) often attributed to ``exoplanets." 
In this discovery paper, we illustrate the power of the squellar model, showing that the light curve of the well-studied ``exoplanet'' WASP-12b can be reconstructed simply from a rotating squar with proportions $1:1/8:1$, without invoking ad-hoc planetary bodies.
Our findings cast serious doubt on the validity of current ``exoplanetary" efforts, which have largely ignored the potential role of squars and have instead blindly accepted the exoplanet hypothesis without sufficient critical scrutiny.
In addition, we discuss the sociopolitical role of climate change in spurring the current exoplanet fervor which has lead to the speculative state of ``exoplanetary science" today.
We strongly urge the astronomical community to take our model proposal seriously and treat its severe ramifications with the utmost urgency to restore rationality to the field of astronomy.
\end{abstract}


\section{Introduction} \label{sec:intro}

The study of exoplanets has emerged as a field of groundbreaking astronomical research, captivating the imaginations of scientists and the public alike. 
Through creative observational techniques and instrumentation, astronomers are ``discovering'' exoplanets at an astounding rate. 
As of early March 2023, there are over 5,300 ``confirmed exoplanets'' in the NASA Exoplanet Archive. 
With the increasing frequency of exoplanet discoveries, we must acknowledge shortcomings in the current status quo. 

The rush to confirm exoplanet discoveries has overlooked potential alternative explanations, leading to a scientific consensus that is overly reliant on untested assumptions and limited data.
In light of this, we aim to encourage critical reflection of current exoplanet discovery methods and to suggest new scientific pathways that may lead us to even more exciting discoveries.

There are currently a variety of exoplanet detection methods, namely (with number of discovered exoplanets): Transit (3970), Radial Velocity (1029), Microlensing (176), Imaging (63), and Other (62). 
The ``transit" method, in which a small and periodic decrease in a star's brightness is assumed to indicate the presence of an additional non-stellar body, singularly accounts for approximately 75\% of all exoplanet detections.
However, it has been shown that some ``transit" light curves can also be explained by other astrophysical phenomena, such as binary stars or sunspots~\citep[e.g.,][]{algol,algol2}.

Rather than exceeding the scope of the evidence and conjuring the existence of a secondary object, we adopt an Occam's Razor approach to variable light curve interpretation. 
Drawing upon extensive knowledge and expertise, we propose a simple theory that at once confidently rejects the existence of ``exoplanets" and expands our current understanding of stellar physics: the existence of cuboid-shape stars, henceforth referred to as squars. 
Our proposal stems from a deep understanding of the complexities of the cosmos, and is well motivated by known naturally occurring cuboid phenomena.

\begin{figure}[tb]
    \includegraphics[width=\columnwidth]{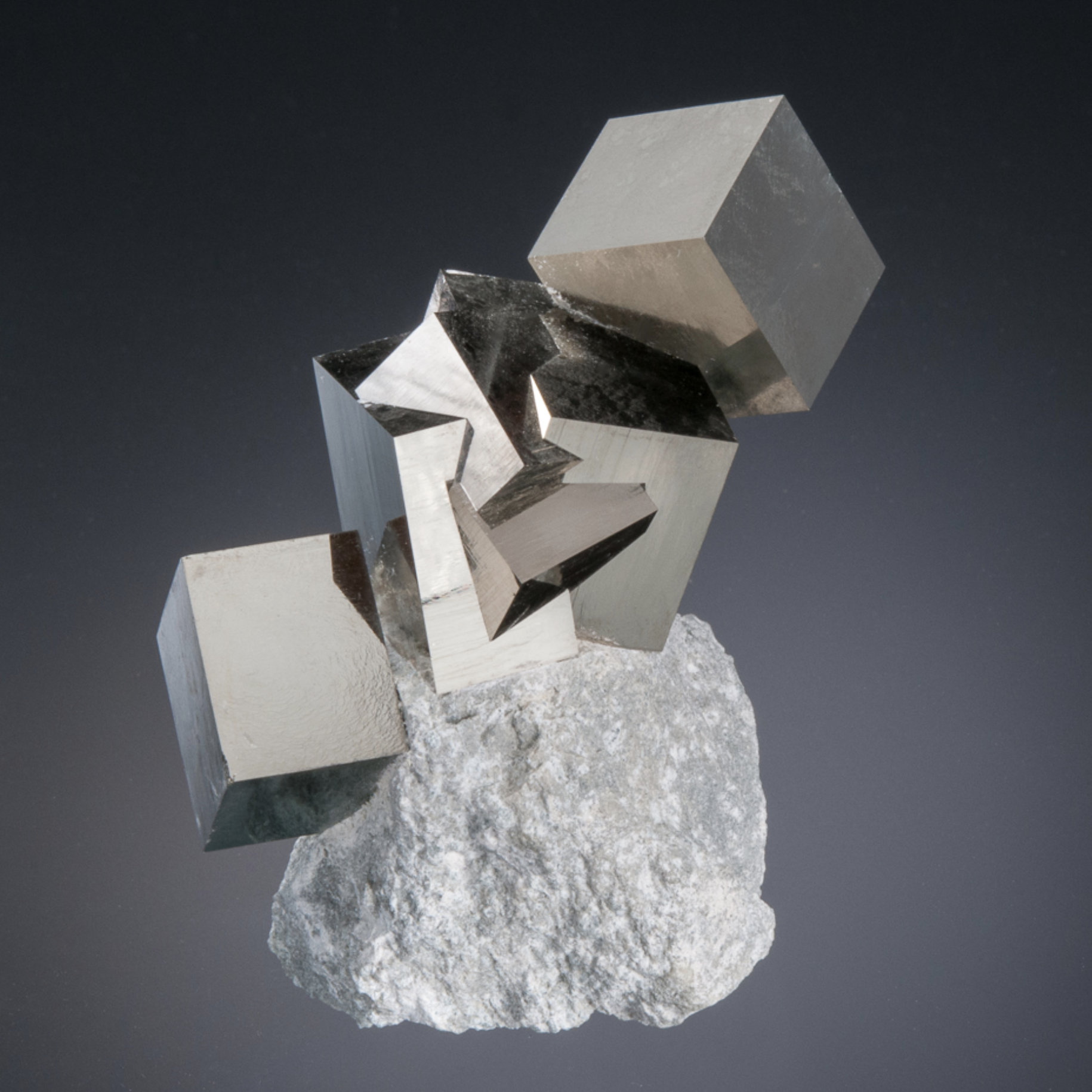}
    \caption{Pyrite Crystals from the Victoria Mine, Navajún, La Rioja, Spain~\citep{pyritecubes}. An example of a perfect cube being found in nature. Top right cube is approximately $7\times7\times7$ cm.}
    \label{fig:pyrite}
\end{figure}

Cuboid structures are ubiquitous in nature, an example of which is shown in Figure~\ref{fig:pyrite}. 
Many minerals, plants, and some bacteria, are known to exhibit cuboidic structure. 
In fact, \citet{Domokos2020} find that natural 3D fragments should have cuboid properties on average, thus interstellar clouds of gas and dust should naturally fragment into the cuboid structure of squars. 

With mother nature as our guide, we present a realistic model for photospheric emission from a rotating squar, thus advancing our understanding of the universe and enabling a transformative paradigm shift for classical astrophysics.
As we dive deeper into the complexities of squellar research, it is imperative that we maintain an open mind and a willingness to challenge our assumptions, including questioning the ubiquitous and quasi-deistic reverence of exoplanet ``research" in the modern age. 
By doing so, we can ensure that our discoveries are based on the most rigorous scientific principles and that our knowledge of the universe continues to expand beyond what could be imagined by lesser scientists.

In Section~\ref{sec:model}, we introduce our novel squellar model and its implications for reproducing observed phenomena in the light curves of stars. 
We fit our model to the existing ``exoplanet" observation of WASP-12b in Section~\ref{sec:tfds}, showing that the data are consistent with a rotating squar. 
In Section~\ref{sec:nonexistence} we present the main conclusion of our work, discuss the scientific ramifications of our discovery, and examine the sociopolitical conditions that lead to the current dogmatic veneration of exoplanets. 
Finally, we discuss our conclusions and present ongoing and potential future work for this exciting field of research in Section~\ref{sec:discussion}.

\begin{figure*}
    \begin{centering}
    \includegraphics[width=0.9\textwidth]{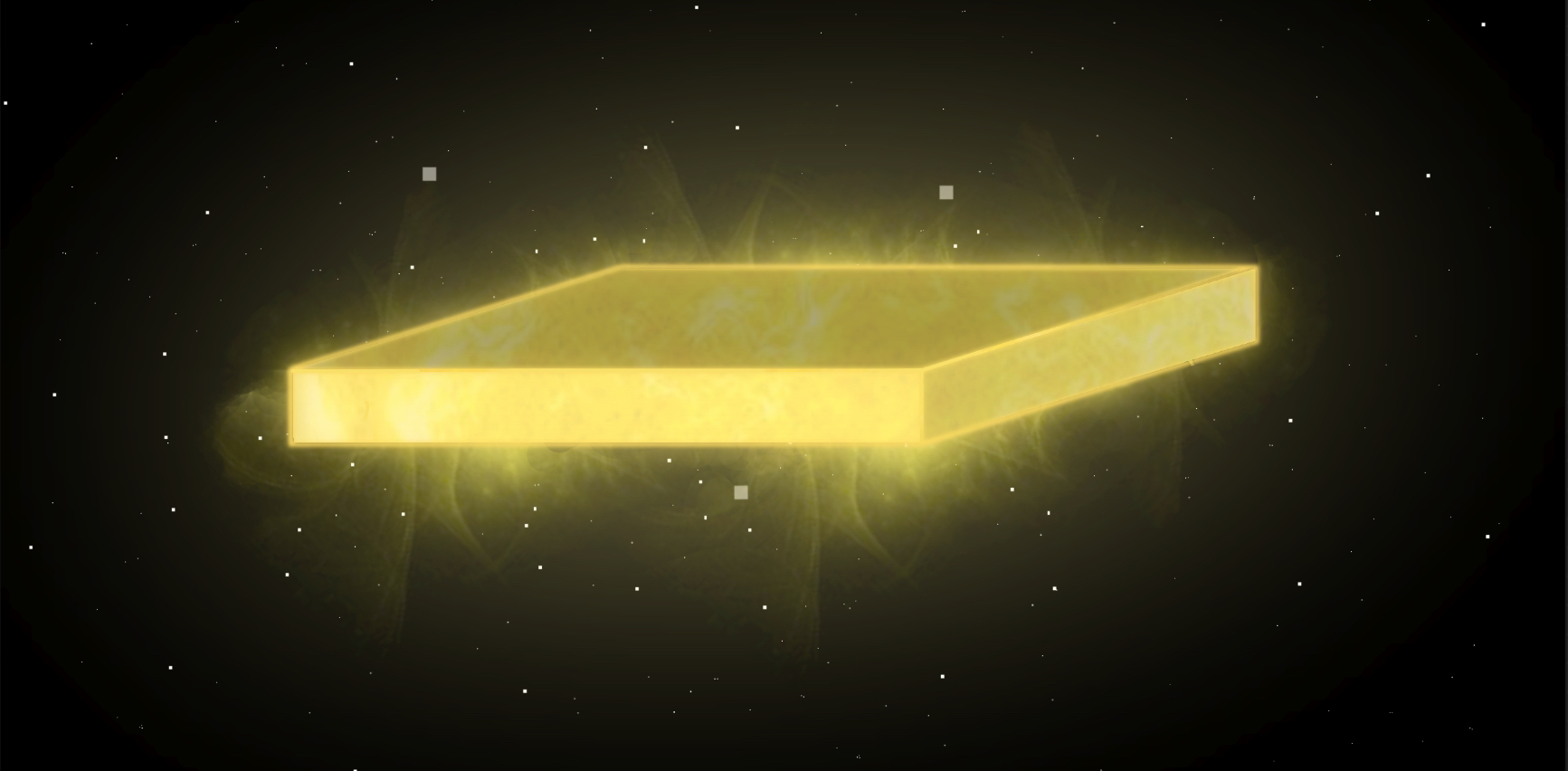}
    \caption{Artist rendition of the squar WASP-12 with physically motivated limb darkening. We note that this squar rotates about one of its long axes. Many of the background stars are also squars, as proposed by our squellar population synthesis models in WHAC Key Paper IV, discussed in Section \ref{subsec:future}.}
    \label{fig:squar}
    \end{centering}
\end{figure*}

\section{Squellar modeling} \label{sec:model}

In this section, we follow in the footsteps of classical stellar models, while implementing an intuitive and physically motivated cuboid geometry.
Traditional stellar evolution codes, such as the open-source stellar evolution software \texttt{Modules for Experiments in Stellar Astrophysics} \citep[MESA;][]{Paxton2011, Paxton2013, Paxton2015, Paxton2018, Paxton2019, Jermyn2023}, are unfortunately insufficient to capture the complexity of geometries extending past 1D. 
We therefore undertake construction of our own comprehensive model of a squar, employing unprecedented state-of-the-art 3D stellar modeling. 
The Woodrum-Hviding-Amaro-Chamberlain model (hereafter the WHAC model) is presented in full in the upcoming WHAC Key Paper I, but is briefly outlined in this section for context.

The WHAC model begins with a generic cuboid that is rotating about one of its axes of symmetry, which can be arbitrarily aligned with respect to a distant observer. 
The projected area of any given face of the squar is given by $A\cos(\beta)$ where $A$ is the inherit area of the squellar face, and $\beta$ is the angle between the normal to the face and the line-of-sight (LOS) to the observer. 
However, the WHAC model's sophistication only begins at the projected area of the squar.

While a naive approach would be to compute the apparent brightness of the star by simply summing the contribution from each visible face of the squar, our model's sophistication enables us to calculate second-order effects, such as limb darkening induced by the squellar atmosphere. 
While typical limb darkening relies on an improvised polynomial in $\cos(\beta)$, the self-consistent magneto-hydrodynamic squellar modeling implemented in WHAC reveals an elegant sigmoid functional form,
\begin{displaymath}
    \ell(\beta,s) =  \left(1 + \exp(s(\beta - 1)) \right)^{-1} = \left({1 + e^{s(\beta - 1)}} \right)^{-1},
\end{displaymath}
where $\ell$ is the proportional reduction in intensity of a given face of the squar and $s$ is the scale parameter that in no way could be considered fine-tuning. 
We have performed a variety of convergence tests to confirm that our squellar model and the resulting limb-darkening prescription is robust to changes in resolution. 
For the remainder of the paper we will adopt a scale parameter of $s=30$, which we consider a natural outcome from our computational model.

Finally, the resultant flux profile generated by a rotating squar cannot be compared to real astrophysical data without detailed consideration of magnetic fields. 
The exclusion of magnetic fields from early models of stellar theory is a notable oversight that we cannot ignore. 
Thus, we have taken great care to incorporate their effects into our own model.
We leave the burden of proof as an exercise to the reader, and simply state that the inclusion of magnetic fields results in an integrated intensity of
    $I_{\rm mag} = I_{\rm no-mag}^{\alpha/2}$,
where $I_{\rm no-mag}$ and $I_{\rm mag}$ are the squellar intensity before and after considering magnetic fields respectively and $\alpha$ is the fine-structure constant.
The use of the fine-structure constant will likely be obvious to any reader equipped with basic squellar astrophysical knowledge.

The WHAC squellar model naturally reproduces trapezoidal flux deviations (TFDs) that are far too commonly attributed to so-called ``exoplanets" in conveniently aligned orbits.
Our modeling framework presents a sophisticated yet elegant resolution to the shapes present in the observed light curves of ``stars" and enables us to fully characterize these objects as cuboid-shaped squars\footnote{In the unlikely case where our WHAC model cannot fully explain a stellar light curve, we have begun to investigate the existence of naturally occurring squellar squatellites or squanets. There are no conclusive results thus far.}.

\begin{figure*}
\centering
\includegraphics[width=0.9\textwidth]{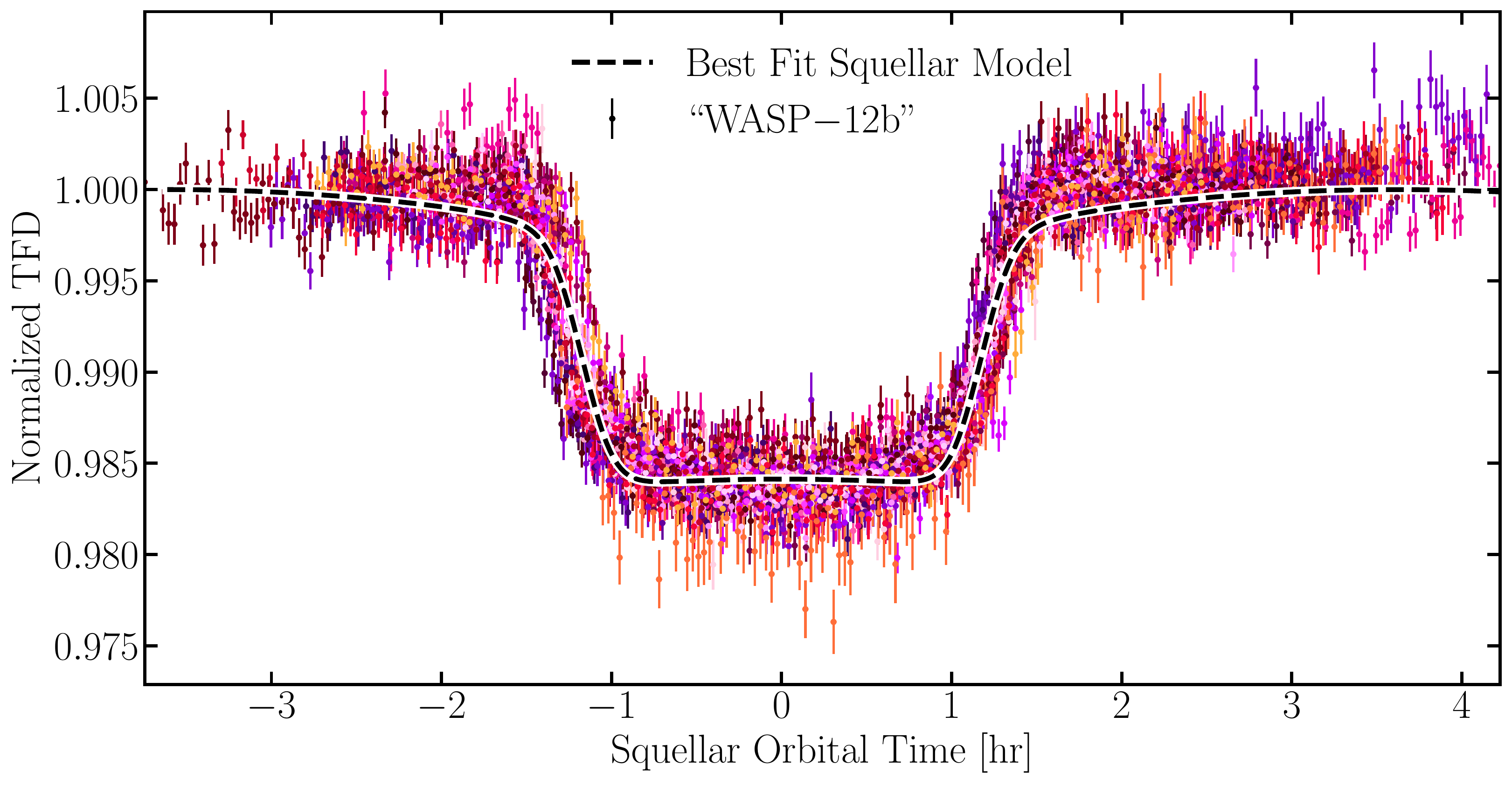}
\caption{Comparison between the best fit squellar model (dashed line) and real light curve data of WASP-12 showing a TFD signature of 1.45\%. Our squellar model is a remarkable fit to the data, thus solidifying our proposal that squars are the natural interpretations of TFDs.}
\label{fig:lightcurve}
\end{figure*}

\section{Fitting Trapezoidal Flux Deviations} \label{sec:tfds}

Our scientific ethos compels us to explore new frontiers and present, for the first time ever, a squellar light curve fit to observed data.
An example of a TFD in real data can be seen in the light curve of 2MASS J063032.79+294020.4 (hereafter WASP-12), identified in broad band optical \citep{Zacharias04} and infrared magnitudes \citep[2MASS]{Skrutskie06-2MASS}. 
WASP-12b, an ``exoplanet'' discovered by \citet{Pollacco06} and announced by \citet{Hebb09-wasp12b}, was invoked to explain the periodic TFD in the light curve of WASP-12. 
Tragically, the literature has never explored the possibility of a squar to explain the light curve of WASP-12. 

It is with great pride that we present the first squellar light curve of a rotating squar. 
We show how simply the flux from a rotating squar perfectly fits the TFD of WASP-12. 
We use 23 light curves\footnote{The light curves are from the Exoplanet Follow-up Observing Program (ExoFOP) website, observed from November 2009 to February 2015 with the Moore Observatory Ritchie-Chretien (MORC) 0.6\,m telescope, which is operated by University of Louisville.} of WASP-12 from \citet{Collins17}. 
Stacking each normalized light curve reveals a characteristic TFD, with a base width approximately 3 hours wide and a total flux deviation of 1.45\%. 
The squellar model has 5 free fit parameters, including the 3 side lengths, which specify the relative dimensions of the squar, and 2 angles, which specify the direction of the angular rotation vector with respect to the line of sight. 
We constrain the 5 free squellar parameters by employing rigorous fitting techniques, including Markov chain Monte Carlo \citep[MCMC;][]{metropolis1953}, to fit the light curve of WASP-12, and find that a squar with relative dimensions of $x = 1$, $y = 1/8$, and $z = 1$ provides the most accurate TFD fit\footnote{The ramifications of one side being a factor of two \textbf{\textit{cubed}} smaller may reveal a deeper significance to the universe, but we do not discuss it further in this work.}. 
Additionally, we find that the squar is inclined towards us at an angle of $\sim$17.2$^\circ$ about its $x$-axis and is rotating about its $z$-axis.
We present an astrophysically-accurate artist rendition of WASP-12 in Figure~\ref{fig:squar}, showcasing the cuboid nature of the squar.

The resultant fit to the light curve created from our best-fit MCMC is presented in Figure~\ref{fig:lightcurve}. 
It is abundantly clear that the WASP-12 TFD signature can effortlessly be recreated with our WHAC squellar model. 
This indisputable fit demonstrates the robust nature of squellar light curve modeling, and implies that all such ``transit" light curves can be easily explained by rotating squellar object rather than ``exoplanets," in stark contrast to the unsubstantiated scientific consensus of the masses. 

In addition, our TFD modeling enables the characterization of additional squellar parameters. 
For example, the period of the TFD signal in WASP-12 is roughly $\sim$26.2 hours, therefore the rotation period of the squar must be approximately 52.4 hours, as we expect to see two symmetric TFDs per single rotation of this particular squellar geometry. 
We also note that the famously changing period of WASP-12 can be neatly explained by introducing well-understood rotational physics to our WHAC squellar model, such as precession and nutation, which we will present in the upcoming WHAC Key Paper II. 

\begin{figure*}
    \begin{centering}
    \includegraphics[width=0.80\textwidth]{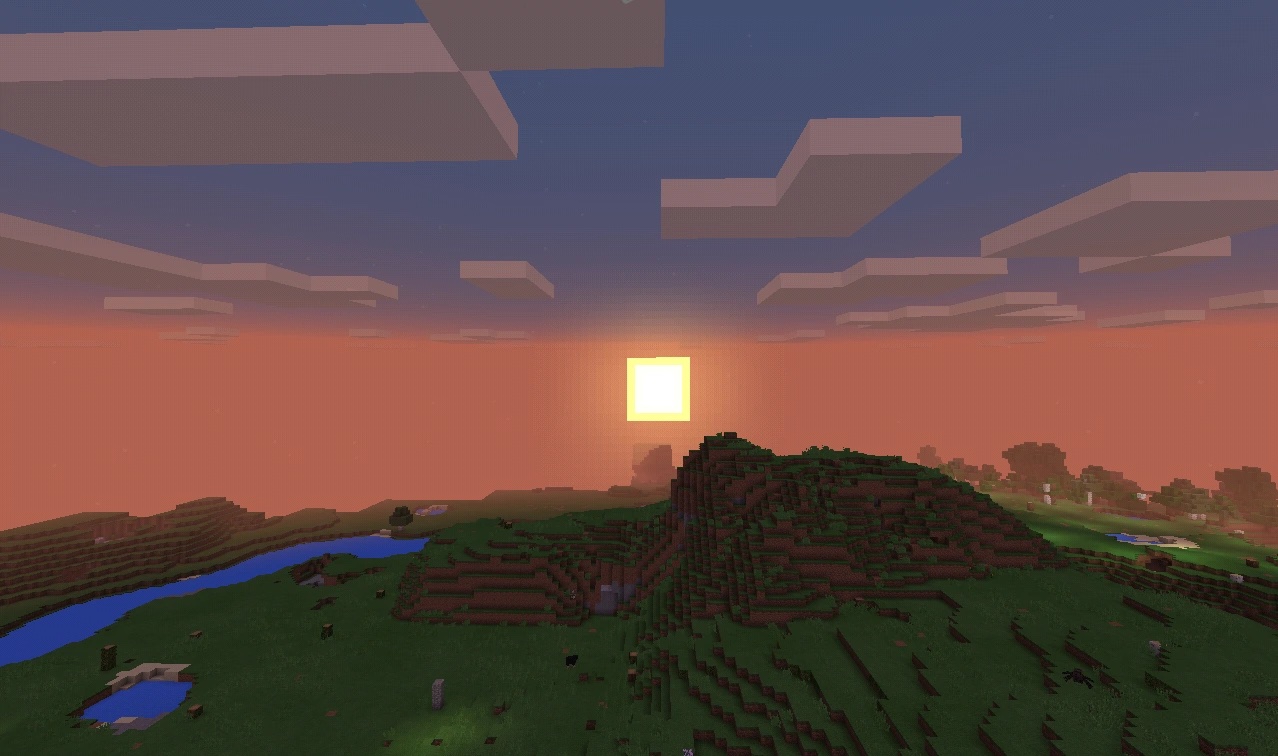}
    \caption{While much of the stellar astrophysics community lags behind in the study of squars, advanced simulations such as Minecraft\textsuperscript{\texttrademark} have already implemented squars into their universe.}
    \label{fig:minecraft}
    \end{centering}
    \end{figure*}

\section{On the non-existence of exoplanets} \label{sec:nonexistence}
Throughout this work, we have shown that the flux signature of a rotating squar naturally produces TFDs. 
It follows that squars are a simpler explanation for the occurrence of TFDs in stellar observations, rather than exoplanets. 
In this section, we discuss the astrophysical implications for the existence of squars and speculate on the sociological and political factors that may have led to the current but invalid astronomical consensus on the existence of exoplanets.

Indeed, the search for the existence of exoplanets likely began as a fear-driven response due to increasing evidence for the existence of climate change. 
In fact, the first Intergovernmental Panel on Climate Change (IPCC) report was published in 1990, and presented evidence that human activities, particularly the burning of fossil fuels, were contributing to global warming and other climate impacts \citep{ipcc1990}.
Unease about the future of life on earth in the face of rising global temperatures and the increased frequency of extreme natural disasters, coupled with post-space-race optimism about off-Earth travel, fueled the exoplanet fiasco. 

In fact, the first discovery of an ``exoplanet" of a solar-mass star occurred just 5 short years later in 1995, with the ``detection" of radial velocity deviations that were concluded to be from a large, Jupiter-mass planet with a period of $\sim$4\,days~\citep{Mayor1995}.
However, a large planet at such small separation from its star would be challenging to explain. 
\cite{Boss1995} showed that large planets, should they exist, will form about 4--5\,AU from stars of roughly 1\,$\Msun$. 
Thus, in order to explain the apparent orbital period, and subsequently, the miniscule orbital radius, complex dynamics such as ``planetary migration" have been invoked. 

The existence of squars offer a much more elegant solution to the problem: exoplanets simply do not exist. 
Unburdened by the need to adhere to the hastily arrived upon scientific consensus, we postulate that \textit{all} ``exoplanet detections" can be explained by the much simpler notion that light curve deviations are due instead to the presence of cuboid squars.
Our modest proposal is radically simpler than existing theories, which are bogged down by unverifiable physical mechanisms,
such as ``planet formation" or the complex dynamics of ``planetary migration."

We reiterate that the quintessential signature of a so-called exoplanet was the nearly trapezoid-shaped flux deviations in the time-series photometry of a star. 
Through a principled approach, our WHAC model has concretely debunked the existence of exoplanets. 

\section{Discussion \& Future Work} \label{sec:discussion}
Ever since the first ``discovery" of an exoplanet, astronomy has been rife with hasty analyses that lead to conclusions that are, frankly, bananas. 
Some authors even claimed that there exist exoplanetary systems that require up to \textit{seven} ``exoplanets" to explain the anomalous light curve data \citep{Gillon16}. 
In this section, we discuss our conclusion on the non-existence of exoplanets and identify avenues and pathways for future work in the exciting new field of squellar astrophysics. 

\subsection{Elegance Over Opulence} \label{subsec:elegance}

Widespread acceptance of exoplanet detections is nothing short of a modern-day Ptolemaic fiasco. 
The constant addition of superfluous orbiting bodies in order to reproduce observed data is clearly reminiscent of the historic heliocentric vs. geocentric tension. 
Before the adoption of the heliocentric model, epicycles were invoked to explain the motions of heavenly bodies under the preconception that the geocentric model must be an immutable truth~\citep{almagest}.
Similarly, exoplanets have been used as the great panacea of the modern age.

Indeed, accepting exoplanets as established truth has enabled the rapid publishing of under-scrutinized and over-hyped scientific hypotheses that only serve to advance fame, glory, and wealth, which are so often associated with the profession of academic astronomy. 
But just as Galileo stood before the church, so do we stand against Big Exoplanet. 

\subsection{The Future of Squellar Astrophysics} \label{subsec:future}

Our rational work has laid the foundation for the introduction of squars to the community, and as a result, an entire avenue of scientific exploration and discovery has been opened. 
In this section we enumerate current efforts being undertaken by our group to expand our understanding of squars and to advance the field of squellar research.

\subsubsection{WHAC Key Paper I:\\SQUABBIT SEASON}
\textit{SQUellar Astrophysicists Bringing Big Ideas Together, SEArching for Squellar mOdels that are New}.
First, we are exploring the possibilities of improving upon our squellar model, such as the inclusion of precession and higher order limb darkening effects. 
While our current squellar model is able to capture the key physical properties of squars, we recognize the benefit of improving our model and collaborating with others who may want to jump ship from the debunked exoplanet field.
Without a proper theoretical framework moving forward, it will be impossible to make progress on our understanding of squars and their significance to galaxy evolution and cosmology. 

It is admittedly shameful that the cutting edge of squellar modeling currently resides in the private sector in the form of a simulation suite that evolves squars in real time. 
This multi-billion dollar code base has hundreds of millions of users, includes extensive modular capabilities, and has been in use for over a decade.
Unfortunately, this pioneering simulation has largely been neglected by stellar researchers.
In Figure~\ref{fig:minecraft}, we present output from the unmodded version of the simulation, known as Minecraft, which depicts a geometrically and astrophysically accurate squar. 

\subsubsection{WHAC Key Paper II:\\Electric Boogaloo}
Next, we extend our analysis to show how other discovery methods for exoplanets can be simply explained by the existence of squars. 
We lay the groundwork for re-evaluating exoplanet data, and invite the now dazed-and-confused exoplanet community to participate in this effort
For example, in WHAC Key Paper II we will demonstrate how squars can explain changes in the radial velocity signature or produce light curve signatures that are consistent with exoplanets ``detected'' via microlensing methods. 

In addition, we are developing extensions of our WHAC model in order to realistically model a neutron degeneracy equation of state, and demonstrate that exoplanets ``detected'' from timing variations can be attributed to cuboid pulsars, or pulsquars.

\subsubsection{WHAC Key Paper III:\\Exoplanets Strike Back$...$ NOT!}
One of the more sensational methods for exoplanet detection involves directly imaging an exoplanet by attempting to block the light from its host star. 
Upon closer evaluation, the small numbers of objects detected in this manner should immediately draw suspicion on their validity. 
In WHAC Key Paper III, we will show that all ``directly imaged exoplanets" are the result of inadequate modeling of the point spread function (PSF) of squars.
The crucial assumption that the star is a spherically symmetric object has considerable consequences for direct detection. 
Despite the vast array of instruments and imaging techniques available, the lack of a proper consideration on the existence of squars implies that the complexity of the PSF has eluded our current modeling techniques, leading to erroneous interpretations of observations.


\subsubsection{WHAC Key Paper IV:\\This is the Skin of a Killer, Bella}

Finally, we bravely endeavor to accurately predict the number of squars present in our universe, which will be presented in the WHAC Key Paper IV.
Trivially, we can set a lower bound on the number of squars as the total number of stars that show TFDs in the largest transit-method surveys.
While not all squars will exhibit TFDs, due to differing viewing angles, the estimate still provides a weak lower limit on the abundance of squars in the Milky Way galaxy.
A full consideration including Bayesian population statistics is currently in progress.

In addition, we will simultaneously put constraints on the number of non-squar, spherical stars in existence in our galaxy, for which we admit there exists at least one.
While some authors claim to have successfully resolved extra-solar stars, we note that improper consideration of the cuboid shape of squars and its interaction with the imaging PSF has likely inflated the number of round stellar shapes, as noted above in our discussion of direct imaging. 







\subsection{Final Thoughts} \label{subsec:thoughts}

We have discovered a new field of astrophysics that not only simplifies our understanding of the universe but also eradicates the need for exoplanets. 
While the implications of this discovery may be disconcerting, it is vital that we remain steadfast in our pursuit of scientific truth. 
We must continue to employ the most rigorous standards of empirical inquiry and theoretical analysis, even as we confront the possibility of significant revisions to our current understanding of the cosmos. 
We are not the first group to go against the status quo \citep[e.g.][]{Gagandeep2022}, and we certainly hope we are not the last.

Therefore, let us remain vigilant in our quest for knowledge, open to the possibility of unexpected discoveries and willing to embrace new insights, however unsettling they may be. 
For it is only through such dedication to scientific rigor that we may achieve a deeper understanding of the universe and our place within it. 


$ $\\ 

KC would like to thank ADHD meds and Target\textsuperscript{\texttrademark}. RA would like to thank her dog, Onyx, for emotional support and ATEEZ (\#8makes1team). 
RA and KC acknowledge emotional support from SKZ-CASE:\#143.
REH acknowledges support from Snuggles the cat and from viewers just like you. 
CAW acknowledges support and inspiration from Dr.~Bug Woodrum.

We would like to thank the MESA team for supporting open software that allowed us to make sense of nonsense, as well as Mojang for their creation of the cutting-edge Minecraft simulations and inspiration of millions of future scientists.

This ``research'' has made use of the NASA Exoplanet Archive, which is operated by the California Institute of Technology, under contract with the National Aeronautics and Space Administration under the Exoplanet Exploration Program. Additionally, this ``research'' has made use of the Exoplanet Follow-up Observation Program (ExoFOP; DOI: 10.26134/ExoFOP5) website, which is operated by the California Institute of Technology, under contract with the National Aeronautics and Space Administration under the Exoplanet Exploration Program. 


\software{\LaTeX\ \citep{lamportLaTeXDocumentPreparation1994}, Matplotlib \citep{hunterMatplotlib2DGraphics2007}, NumPy \citep{oliphantGuideNumPy2006,vanderwaltNumPyArrayStructure2011, harrisArrayProgrammingNumPy2020}, WHAC (WHAC Key Paper I)}

\bibliography{squarsrefs}{}
\bibliographystyle{aasjournal}

\end{document}